\documentclass[aps,prl,twocolumn,showpacs,groupedaddress,superscriptaddress]{revtex4-1}
\usepackage{amsmath,amssymb,graphicx}

\def\phiPRP {\phi_{\rm PRP}}

\begin{document}


\title{Excess optical quantum noise in atomic sensors}

\author{Irina Novikova}
\email[ixnovi@wm.edu]{}
\affiliation{College of William \& Mary, Williamsburg, VA, 23185, USA}

\author{Eugeniy E. Mikhailov}
\affiliation{College of William \& Mary, Williamsburg, VA, 23185, USA}

\author{Yanhong Xiao}
\affiliation{Fudan University, Shanghai, China}

\date{\today}

\begin{abstract}

Enhanced nonlinear optical response of a coherent atomic medium is the basis for many atomic sensors, and their performance is ultimately limited by the quantum fluctuations of the optical read-out. Here we demonstrate that off-resonant interactions can significantly modify the quantum noise of the optical field, even when their effect on the mean signal is negligible. We illustrate this concept by using an atomic magnetometer based on the nonlinear Faraday effect: the rotation of the light polarization is mainly determined by the resonant light-induced spin alignment, which alone does not change the photon statistics of the optical probe. Yet, we found that the minimum noise of output polarization rotation measurements is above the expected shot noise limit.  This excess quantum noise is due to off-resonant coupling and grows with atomic density. We also show that  the detection scheme can be modified to reduce the measured quantum noise (even below the shot-noise limit) but only at the expense of the reduced rotational sensitivity.  These results show the existence of previously unnoticed factors in fundamental limitations in atomic magnetometry and could have impacts in many other atom-light based precision measurements.

\end{abstract}


\maketitle

\begin{figure}
 \includegraphics[width=1.0\columnwidth]{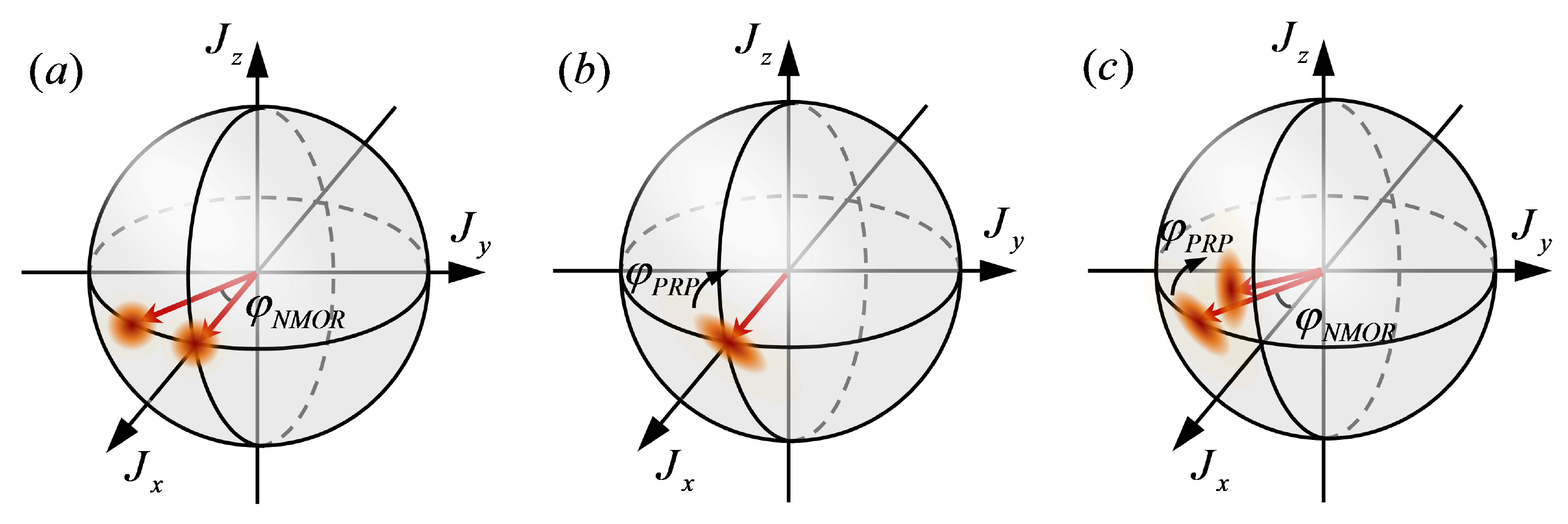}%
 \caption{\label{fig:cartoon}
 (a) Polarization rotation of a linearly polarized coherent optical field under a magnetic field, depicted in a Poincar\'{e} sphere. The initial linear polarization is along $J_x$, and the balanced polarimeter detection measures the value of $J_y$. Quantum uncertainty in the Stokes vector direction is represented by the noise ellipse (or circle, for a coherent state).  (b) Output polarization of a linearly polarized light taking into account of off-resonant atom-light interaction without a magnetic field. The total Hamiltonian has a shearing effect on the uncertainty circle, turning it into an ellipse, resulting in a squeezed state of light. By adjusting the phase between $x$ and $y$ polarization component ($\phiPRP$), one can rotate the noise ellipse, making possible a sub-shot noise detection of $J_y$.  (c) When a small magnetic field is applied, the noise in the magnetic field measurement (proportional to $J_y$ fluctuations) is above the shot-noise level, since the long axis of the noise ellipse is almost along the equator of the Poincar\'{e} sphere.  Rotation of the Stokes vector around the $J_x$-axis by $\phiPRP$, as in (b), can decrease noise but at the price of a reduced $J_y$ signal value.}
 \end{figure}

Atoms are nature's most sensitive measurement devices, and their strong optical responses to external fields enable both precision measurements of fundamental constants as well as realization of  practical devices for detection of electric and magnetic fields~\cite{adv_atom_phys_book}. Since rotation of optical polarization often serves as the measurable output~\cite{budker_optmagn_book}, the sensitivity of such sensors is ultimately limited by quantum polarization fluctuations (the  photon shot noise), as illustrated in Fig.~\ref{fig:cartoon}(a), and can be surpassed only  if the input optical field is squeezed~\cite{adv_atom_phys_book, mitchel2010prl_sqz,mikhailov2012sq_magnetometer}.

It is generally assumed that light-atom interactions do not change photon statistics, and that measurements remain shot noise-limited. 
In this letter, we demonstrate that this assumption is not always true, and that unaccounted interactions with off-resonant atomic levels can lead to excess quantum noise, even if their effect on the mean signal remains negligible. 
Our findings identify an additional fundamental limitation for optical measurements involving atoms, and should have a significant impact for a wide range of atomic sensors.

  We illustrate the origin and impact of the excess quantum noise in an atomic magnetometer, based on the nonlinear magneto-optical (Faraday) rotation (NMOR) effect.  In such a device, an external magnetic field causes a rotation of linear polarization in the vicinity of atomic resonances.  The nonlinear magneto-optical response, based on the light-induced alignment of atomic spins, surpasses the linear response by several orders of magnitude due to the long spin coherence lifetime~\cite{budkerRMP02}.
 There is no mean polarization rotation without a magnetic field; yet, the coupling of light to an off-resonant excited level modifies the quantum fluctuations of the optical field~\cite{matsko_vacuum_2002,hsu_effect_2006}. Such nonlinear interaction has been used to produce squeezed vacuum in orthogonal polarization of the optical field, known as polarization self-rotation (PSR) squeezing~\cite{matsko_vacuum_2002,ries_experimental_2003,mikhailov2008ol,grangier2010oe} or polarization squeezing~\cite{giacobino2003prl,lezama2011pra} (see Fig.~\ref{fig:cartoon}(b)). In this work, we consider the effect of this off-resonant interaction on  polarization rotation measurements due to the nonlinear Faraday effect. 
Our previous studies~\cite{HorromJPB12} have shown that the PSR squeezing persists even if a small non-zero  longitudinal magnetic field is applied. We found, however, that the measurements of the resulting polarization rotation are accompanied by the excess quantum noise above the shot noise level. Sub-shot noise measurements correspond to reduced magnetic field sensitivity, as illustrated in Fig.~\ref{fig:cartoon}(c) and explained later in the paper.

\begin{figure}
 \includegraphics[width=1.0\columnwidth]{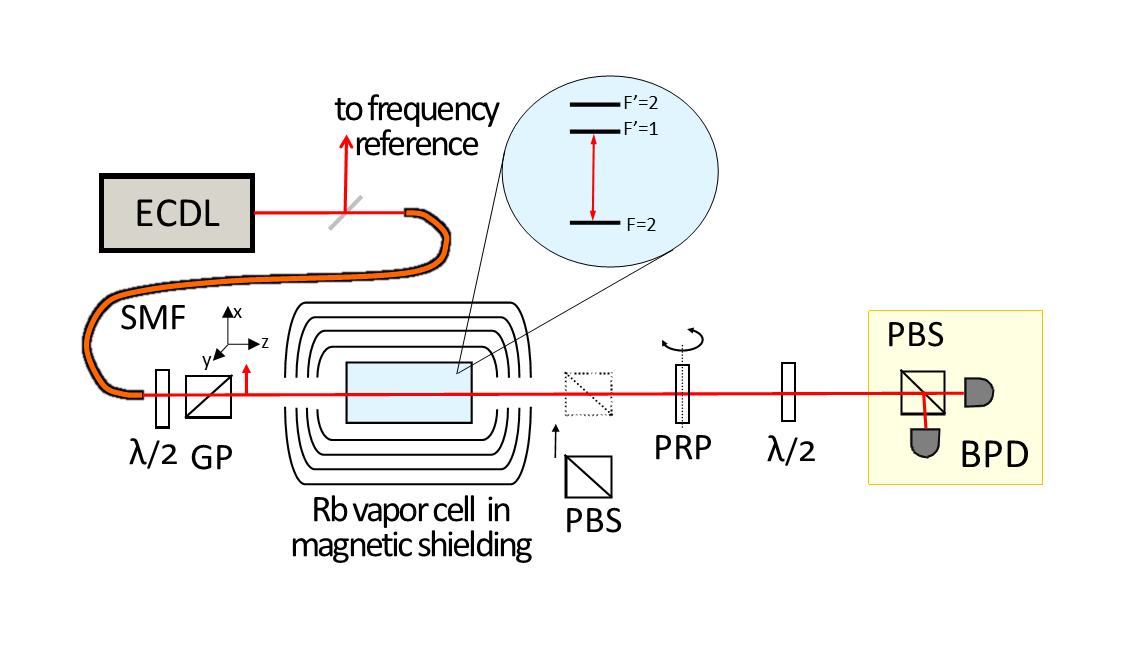}%
 \caption{\label{fig:setup}
 Schematics of the experimental setup (see text for abbreviations). Inset shows the detuning of the laser with respect to Rb atomic levels. }
 \end{figure}

The schematic of the experiment is shown in Fig.~\ref{fig:setup}. 
We used a paraffin-coated cylindrical Pyrex cell (75~mm in length,  25~mm in diameter), containing isotopically enriched $^{87}$Rb vapor. The cell was mounted inside a four-layer magnetic shielding, and the number density of Rb atoms was controlled by adjusting the temperature of the pull-off tip of the cell. An external cavity diode laser (ECDL) was tuned to the $5^{2}S_{1/2} F=2\rightarrow 5^{2}P_{1/2},F^{\prime} = 1$ transition of the ${}^{87}$Rb ($\lambda \simeq$ 795~nm). The laser output passed through a single-mode optical fiber (SMF) followed by a Glan-laser polarizer (GP)  to prepare a high-quality linearly-polarized collimated beam with a diameter of approximately $2.3$~mm. The input laser power in the cell was controlled by rotating a half wave plate before GP, with maximum available power $24$~mW. The polarization rotation and the quantum noise of the output optical field were analyzed by rotating its polarization by $45^\circ$ with a half-wave plate ($\lambda/2$) with respect to the axes of a polarizing beam splitter (PBS), and then by sending the PBS outputs to a balanced photodetector (BPD) with $1.6\times 10^{4}$~V/A gain, 9~MHz 3~dB bandwidth, and dark noise level at least 10~dB below the shot noise level. 

The strong linearly polarized field played a role of the local oscillator (LO) for quantum noise measurements of the PSR squeezing. The relative phase between the two polarizations (i.e. between the local oscillator and the analyzed vacuum field) was adjusted by  horizontally tilting a phase-retarding plate (PRP)---a half-wave plate with optical axes aligned with the LO. The shot noise level measurements were done with a polarizing beam splitter (PBS) placed after the Rb cell, transmitting the LO only.

\begin{figure}

 \includegraphics[width=\columnwidth]{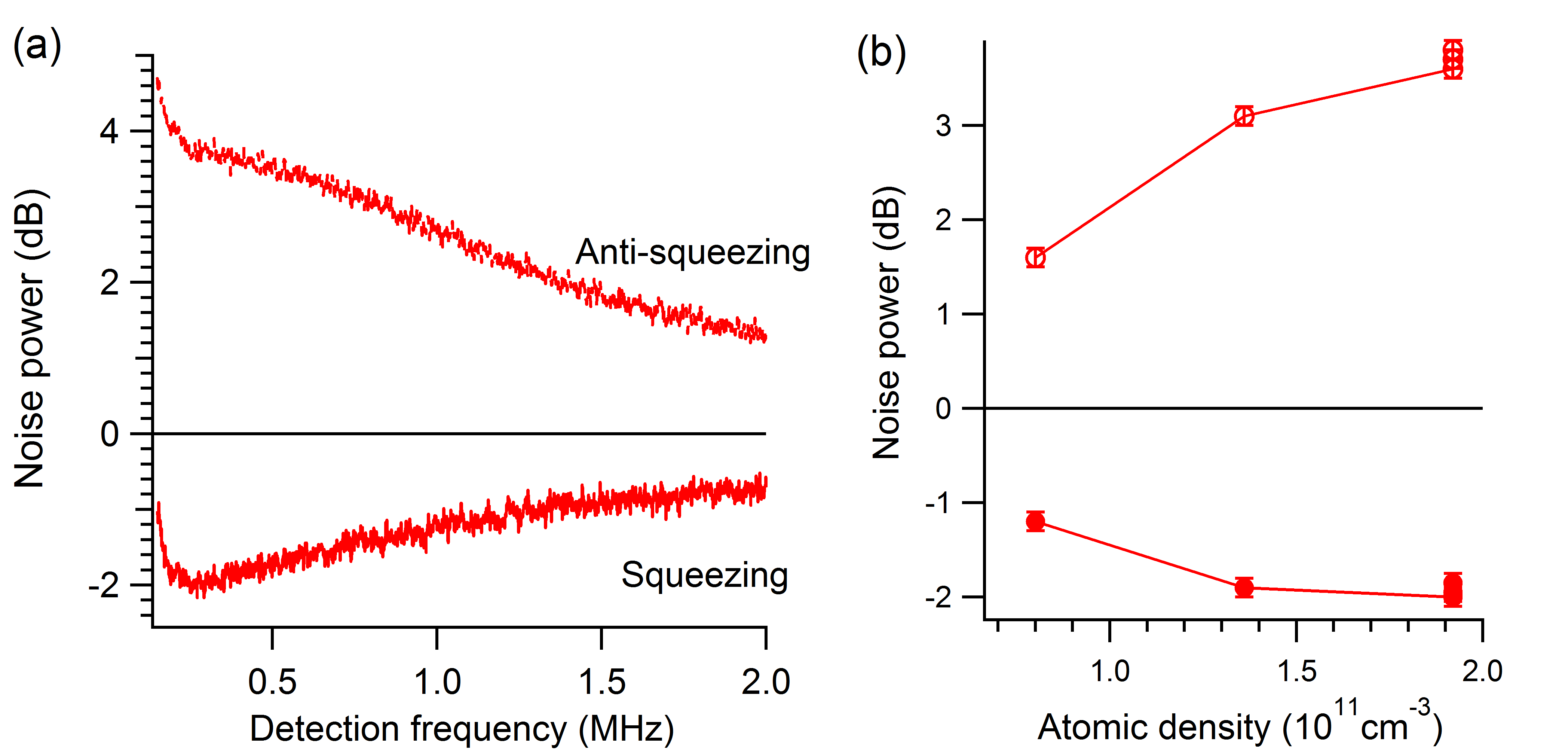}%
 \caption{
	\label{fig:sqzsamples}
	\emph{(a)} Minimum and maximum measured quadrature noise spectra as a
	function of detection frequency. \emph{(b)} Atomic density dependence of
	the squeezed and anti-squeezed quadratures. The incident laser power is
	$P=4$~mW. The noise power is normalized to the shot noise level (0~dB
	level), the spectrum analyzer settings are RBW=30~kHz, VBW=30~Hz. The
	uncertainties are one standard deviation.
}
\end{figure}

We found that the overall quantum noise spectra, shown in Fig.~\ref{fig:sqzsamples}(a), were similar to those previously reported~\cite{ries_experimental_2003,mikhailov2008ol,grangier2010oe,lezama2011pra}. However, the presence of the paraffin coating significantly improved the purity of the output squeezed state (\textit{i.e.}, the difference between maximum squeezing and maximum anti-squeezing with respect to the shot noise level), mainly because of much longer ground state coherence lifetime ($>\!\!100$~ms here). Previous measurements, performed in uncoated vapor cells, reported squeezing of $2-3$~dB below the standard quantum limit (SQL) with the corresponding anti-squeezed quadrature noise to be $10-20$~dB above the SQL~\cite{ries_experimental_2003,mikhailov2008ol,grangier2010oe,lezama2011pra}. Fig.~\ref{fig:sqzsamples}(b) shows a roughly linear growth of both squeezed and anti-squeezed quadratures with the number density of Rb atoms.  The highest value of squeezing---approximately $2.0$~dB below SQL---was obtained at the highest safe operational temperature (54$^\circ$C) of the paraffin-coated cell. The corresponding anti-squeezing quadrature noise was $3.5-4.0$~dB above SQL.

For the NMOR measurements, the longitudinal magnetic field was produced by a solenoid mounted inside the magnetic shielding. The polarization rotation angle was measured using a balanced polarimeter  with zero phase between two polarizations (equivalent to no phase-retarding plate). We refer to the quantum noise,  measured at these conditions, as \emph{intensity} quadrature of the output light. The \emph{phase} quadrature was measured with the PRP phase changed by $90^\circ$. The \emph{squeezed} and \emph{anti-squeezed} quadratures corresponded to the minimum and maximum quantum noise values, obtained when the PRP phase was adjusted.

\begin{figure}
 \includegraphics[width=0.8\columnwidth]{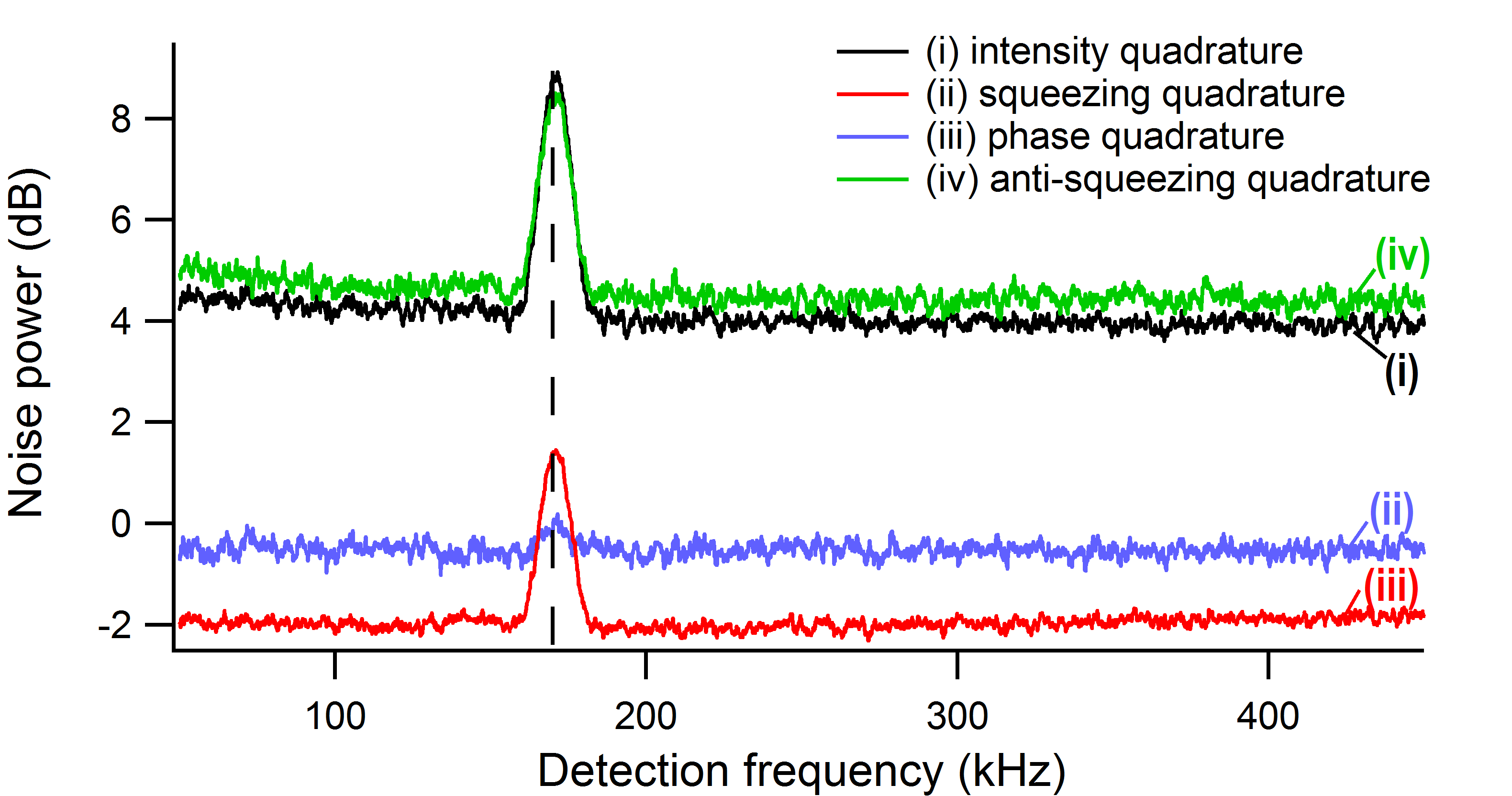}%
 \caption{\label{fig:sqzquad}
 Nonlinear Faraday rotation and noise power measurements for various quadratures. Magnetic field is modulated at $170$~kHz (vertical dashed line). Individual curves \emph{(i)-(iv)} are recorded at fixed positions of the phase-retarding plate $\phiPRP$. The incident laser power is $P= 4$~mW, the spectrum analyzer settings are RBW=30~kHz, VBW=30~Hz. The noise power is normalized to the shot noise level, set to be 0~dB.}
 \end{figure}

To observe the relationship between the NMOR signal and the measured quantum noise, we applied a modulated magnetic field in the Rb cell, and monitored the strength of the signal and noise for different tilt positions of the PRP (thus changing the phase $\phiPRP$ between the two light polarizations).  
Figure~\ref{fig:sqzquad} 
 clearly shows that the maximum NMOR signal was observed for $\phiPRP=0$.  At the same time, the measured noise background was above the shot-noise limit, with the measured noise nearly equal to the maximum anti-squeezing quadrature noise.
As the phase-retarding plate was rotated to access different noise quadratures, the NMOR signal decreased gradually to zero for $\phiPRP=90^\circ$, which corresponded to phase noise quadrature measurements, where the measured noise was nearly at the shot noise level.  
Achieving the maximum noise suppression (squeezing quadrature) required additional small $\phiPRP$ adjustment, and even though some NMOR signal was detectable, it was strongly suppressed, compared to the intensity quadrature measurements.

This analysis clearly shows that the squeezing and anti-squeezing noise quadratures did not perfectly match the intensity or phase quadratures, but that the squeezed noise ellipse was rotated by a small angle with respect to the intensity quadrature, as shown in Fig.~\ref{fig:cartoon}(b), giving rise to the elevated noise level in the polarization rotation measurements. These measurements are in qualitative agreement with 
phenomenological PSR squeezing~\cite{matsko_vacuum_2002}, which predicted  excess noise in the intensity quadrature, shot noise level at the phase quadrature, and non-zero squeezing angle for squeezed and anti-squeezed quadratures.

We can gain some intuitive understanding of the squeezing process by employing a Poincar\'{e} sphere analysis of light polarization, shown in Fig.~\ref{fig:cartoon}.  We use a simplified level system to describe Rb atoms interacting with a monochromatic optical field, shown in the inset of the Fig.~\ref{fig:setup}. In this system, the atom-light interaction part of the Hamiltonian has two parts, corresponding to resonant and off-resonant interactions with the Zeeman levels within the two hyperfine excited states~\cite{sherson2006thesis}:

\begin{equation} \label{hamiltonian}
\hat{H}_{int} = A  \hat{J}_z \cdot \hat{S}_z + \hat{H}_{EIT}.
\end{equation}
The first term describes the off-resonant interaction of light with the atomic spin in the $z$-direction $\hat{S}_z$, mediated by the multiple upper excited states. Here, we neglect the residual absorption from that excited state.
The second term describes the resonant EIT-like interaction, which is responsible
for optically pumping the atoms into the non-interacting dark
state~\cite{FleischhauerRevModPhys05,novikova2012review}. As a result,
the atomic population distribution, and thus the mean atomic spin value,
is determined by the differences in $E_\pm$, the circular components of
the original optical field: $S_z \propto J_z \propto{E_+^2-E_-^2}$, and
thus the quantum fluctuations in $J_z$ cause fluctuations in $S_z$. In
this case, the off-resonant interaction term in Eq.(\ref{hamiltonian})
becomes proportional to $J_z^2$, which shears the originally symmetric
ball of uncertainty into an ellipse with a long axis tilted towards the
equator, as illustrated in Fig.~\ref{fig:cartoon}(c). Interestingly, the
$J_z^2$ interaction is analogous to the one-axis twisting Hamiltonian
$S_z^2$ used to obtain spin-squeezed atomic ensemble~\cite{spin-squeeze93,
VuleticPRL2010}. As discussed in~\cite{VuleticPRL2010}, the shearing strength is
proportional to atomic density, which in trend agrees with the density
dependence of the squeezing and anti-squeezing quadrature shown in
Fig.~\ref{fig:sqzsamples}(b).

The PRP tilting leads to a rotation of the noise ellipse around the $J_x$ direction, and thus the noise ellipse has to be rotated almost by $90^\circ$ to reach the minimum noise quadrature. 
Polarization rotation due to the applied magnetic field rotates the Stokes vector along the equator of the Poincar\'{e} sphere, and thus almost along the elongated axis of the noise ellipse, leading to excess detection noise described above. At the same time, any change in PRP phase results in the reduction of the detected rotation signal, as the Stokes vector rotates out of the equatorial plane, as shown in Fig.~\ref{fig:cartoon}(c). This shows that the maximum response to the magnetic field is always accompanied by increased quantum noise. It may be possible still to take advantage of PSR squeezing, if the orientation of the PRP could be dynamically locked to the mean polarization direction, so that the noise ellipse can be rotated independently. Alternatively, one can use the polarization squeezed light or twin-beam detection to decrease the detection noise floor~\cite{mitchel2010prl_sqz, mikhailov2012sq_magnetometer,otterstrom2014SelfSqMag}. However, realization of such methods in practice is rather technically involved.

\begin{figure}
 \includegraphics[width=1.0\columnwidth]{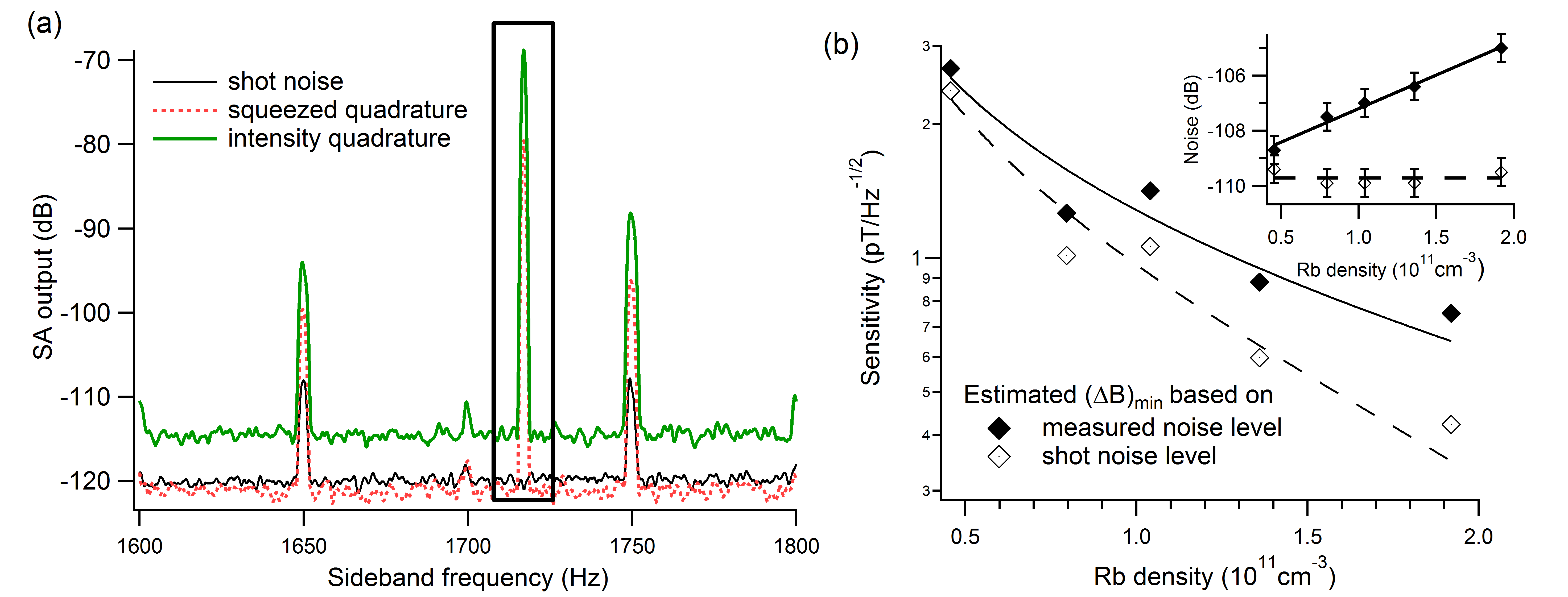}%
 \caption{\label{fig:magnsens}
 \emph{(a)} Nonlinear Faraday rotation signal for a sinusoidally-modulated magnetic field ($\mathrm{f}_{mod}= 1,717$~Hz), measured in the intensity quadrature ($\phiPRP=0$); in the squeezing quadrature ($\phiPRP$ is set to minimize the quantum noise);  the measured shot noise level is also shown. The relevant signal is in the box, two neighboring peaks are due to technical noise.  For these measurements the spectrum analyzer settings are RBW=VBW=1~Hz.
 \emph{(b)}  Measured quantum noise limited (solid diamonds) and estimated shot-noise limited (hollow diamonds) magnetic field sensitivity as functions of atomic density. \emph{Inset:} Measured intensity quadrature noise and shot noise as a function of atomic density. The lines are to guide the eyes.}
 \end{figure}

The observation of additional intensity noise via the PSR interaction may have serious consequences for atomic magnetometers or other devices that are based on nonlinear polarization rotation. Typically, their performance improves with atomic density due to collective enhancement of coherent light-atom interactions. However, the detection intensity noise also grows with atomic density, because the shearing factor of the uncertainty ellipse for the Stokes vector is proportional to atomic density. This will have an important impact on magnetometer performance. For the wide range of laser powers and magnetic field parameters we explored, the intensity noise quadrature, corresponding to the maximum rotation signal, was always above the shot-noise level, as shown in Fig.~\ref{fig:magnsens}(a). We verified that the signal-to-noise ration cannot be further improved by adjustments of $\phiPRP$. For example, when the detection scheme was adjusted to the minimum quantum noise (squeezed quadrature), the measured value of polarization rotation dropped by approximately $10$~dB. The measured intensity quadrature noise increased linearly with atomic density (see inset in Fig.~\ref{fig:magnsens}(b)).

We can now evaluate the dependence of the magnetic field measurement sensitivity as a function of atomic density and analyze the effect of the elevated intensity noise.
We estimated the signal-to-noise of such a ``magnetometer'' by measuring the polarization rotation response to the modulated applied magnetic field and the corresponding noise floor (with the PRP removed) for the range of atomic densities. We also recorded the shot noise level, and estimated both realistic and shot-noise-limited values for the minimum detectable magnetic field, as shown in Fig.~\ref{fig:magnsens}(b). It is clear that even though the measured magnetic field sensitivity of such a device continues to improve with the growth of the atomic density, realistically it improves at lower rate than estimated only by the shot noise limited sensitivity.

In conclusion, we have investigated the relationship between nonlinear magneto-optical rotation and polarization self-rotation squeezing to optimize polarization rotation measurements in Rb vapor. We found that the noise ellipse, resulting from PSR squeezing, is tilted by a small angle with respect to the intensity quadrature, resulting in excess detection noise when the polarization rotation is measured. Both rotation signal and the excess noise increase with the atomic density, making this effect more pronounced in the region where maximum rotational sensitivity is expected. This analysis demonstrates the importance of enhanced off-resonant interactions that can affect the quantum fluctuations in dense coherent atomic ensembles.

IN acknowledges the support of AFOSR grant FA9550-13-1-0098 and NSF grant PHY-1308281. YX acknowledges the support from NBRPC(973 Program Grants No. 2012CB921604 and 2011CB921604), NNSFC (Grants No. 11322436), and the Research Fund for the Doctoral Program of Higher Education of China. We thank N.~B. Phillips for the help with manuscript preparation.

\end{document}